\documentclass[useAMS,usenatbib]{mn2e}

\usepackage{graphicx}
\usepackage{amssymb}

\newcommand{\new}[1]{{\rm #1}}

\title[Testing Newtonian Gravity With Globular Clusters]{Testing
  Newtonian Gravity with AAOmega:\\Mass-to-Light Profiles of Four
  Globular Clusters}
\author[Richard R. Lane {\it et al.}]{Richard R. Lane$^{1}$\thanks{E-mail:
rlane@physics.usyd.edu.au}, L\'aszl\'o L. Kiss${^1}$, Geraint
  F. Lewis${^1}$, Rodrigo A. Ibata${^2}$,\vspace{2.5mm}\\
\hspace{-1mm}{\LARGE {\rm Arnaud Siebert${^2}$, Timothy
    R. Bedding${^1}$ and P\'eter Sz\'ekely$^{3}$}}\\
$^{1}$Sydney Institute for Astronomy, School of Physics, A28,
  The University of Sydney, NSW, Australia 2006\\
$^{2}$Observatoire Astronomique, Universite de Strasbourg, CNRS,
  67000 Strasbourg, France\\
$^{3}$Department of Experimental Physics, University of Szeged, Szeged 6720,
Hungary\\
}
\begin{document}

\date{This draft: \today}

\pagerange{\pageref{firstpage}--\pageref{lastpage}} \pubyear{2009}

\maketitle

\label{firstpage}

\begin{abstract}
Testing Newtonian gravity in  the weak-acceleration regime is vital to
our understanding  of the nature of the  gravitational interaction. It
has  recently been claimed  that the  velocity dispersion  profiles of
several globular  clusters flatten out at large  radii, reminiscent of
galaxy rotation  curves, even though globular clusters  are thought to
contain little  or no dark  matter.  We investigate this  claim, using
AAOmega observations  of four globular clusters, namely  M22, M30, M53
and M68.  M30, one such cluster that  has had this claim  made for its
velocity  dispersion,  was   included  for  comparison  with  previous
studies.   We  find no  statistically  significant  flattening of  the
velocity dispersion at large radii  for any of our target clusters and
therefore we infer the observed  dynamics do not require that globular
clusters  are dark  matter dominated,  or a  modification  of gravity.
Furthermore,  by applying a  simple dynamical  model we  determine the
radial  mass-to-light  profiles  for  each  cluster.   The  isothermal
rotations of each cluster are also measured, with M22 exhibiting clear
rotation, M68 possible rotation and  M30 and M53 lacking any rotation,
within the uncertainties.
\end{abstract}


\section{Introduction}

Newtonian gravitation has been shown to describe the motions of bodies
with  intermediate  accelerations  (e.g.   Solar System  bodies)  very
accurately.   The  Newtonian theory  of  gravity  breaks  down in  the
presence   of  strong   gravitational  fields,   where  it   has  been
successfully superseded  by General Relativity, but  there is evidence
that  there are  also  discrepancies in  the low-acceleration  regime.
Spiral galaxies,  for example, require  the invocation of  Dark Matter
(DM) to reconcile the  discrepancy between their rotational velocities
and our understanding of Newtonian gravity.  Another indication is the
so-called ``Pioneer  anomaly'': radiometric data from  Pioneers 10 and
11  have shown  that  both are  experiencing  an unexplained  constant
acceleration  of  $a=(8.74\pm1.33)\times10^{-10}$\,m\,s$^{-2}$  toward
the Sun  \cite[see][and references therein]{deDiego08}.   In fact, all
spacecraft  in the  outer Solar  System seem  to be  experiencing this
anomalous   acceleration  (e.g.    \citealt{Anderson02}),  potentially
indicating a deviation from Newtonian gravity at low accelerations.

It  has been  claimed that  one of  the leading  versions  of modified
Newtonian dynamics \cite[MOND;][]{Milgrom83} can predict the kinematic
properties of galaxies without  invoking DM. MOND predicts a breakdown
of     Newtonian     gravity      at     an     acceleration     scale
$a_0\approx1.2\times10^{-10}$\,m\,s$^{-2}$  \cite[see][and  references
therein]{Iorio09}. This is approximately the acceleration regime where
DM  becomes necessary  to reconcile  theoretical velocity  profiles of
galaxies with observation.

Globular clusters  (GCs) are  the ideal test  bed for  this divergence
from Newtonian  predictions, since they are thought  to contain little
or  no  DM.    This  is  based  on  evidence   from  dynamical  models
\cite[e.g.][]{Phinney93},        {\it        N}-body       simulations
\cite[e.g.][]{Moore96},    observations     of    GC    tidal    tails
\cite[e.g.][]{Odenkirchen01},  dynamical and  luminous  masses of  GCs
\cite[e.g.][]{Mandushev91}  and the lack  of microlensing  events from
GC-mass  dark haloes  \cite[e.g.][]{Navarro97,Ibata02}.   In addition,
GCs  at  different  distances  from  the  Galactic  centre  experience
differing  gravitational  attractions from  the  Galaxy,  so that  any
Galactic  influence   can  be  ruled   out  if  all   exhibit  similar
behaviour. Furthermore, the accelerations  experienced by stars in GCs
drop below $a_0$ well inside the tidal radius \cite[e.g.][]{Scarpa07}.

Recently, \cite{Scarpa07} showed that there may be a flattening of the
velocity  profiles  of several  GCs  in  the  Galactic halo  ($\omega$
Centauri, M15, M30, M92 and  M107), at or about the acceleration where
DM  is invoked  to maintain  stability in  rotating galaxies.   On the
other hand, \cite{Moffat08}  overlaid the \cite{Scarpa07} results with
a Modified Gravity (MOG) model and found little, or no, deviation from
Newtonian  gravity,  for  GCs  with  masses  less  than  a  few  times
$10^6$\,M$_\odot$.   Because  both groups  used  an  identical set  of
measured velocity  dispersions, there  is an obvious  discrepancy.  If
MOND were  correct, it would mean that  all gravitational interactions
would  diverge from  Newtonian  gravitation in  the predicted  regime.
Irrespective of MOND itself,  testing the gravitational interaction at
low accelerations is extremely  important in the overall understanding
of gravity.  This  paper continues this test, with  four Galactic halo
GCs, namely  M22 (NGC 6656),  M30 (NGC 7099),  M53 (NGC 5024)  and M68
(NGC 4590).

\section[]{Data Acquisition and Reduction}\label{Data Aquisition and
  Reduction}

We used  the multi-object,  double-beam spectrograph (AAOmega)  on the
3.9\,m Anglo-Australian Telescope at  Siding Spring Observatory in New
South  Wales,  Australia,  to  obtain  the spectra  for  this  survey.
AAOmega has 392  fibres covering a 2 square degree  field of view with
each fibre capable  of obtaining a single spectrum  from a single star
in  this  field.   The   1500V  grating  \new{(with  a  resolution  of
  $R=3700$)}  was  used  in  the   blue  arm  and  the  1700D  grating
\new{($R=10000$)}  was   used  in  the  red  arm,   with  the  central
wavelengths   set  to  5200\AA   ~and  8640\AA,   respectively.   This
configuration  was chosen  to  include the  calcium  triplet lines  at
8498\AA, 8542\AA ~and 8662\AA, for accurate velocity determination and
proxy metallicities, as well as the swathe of iron and magnesium lines
around   5200\AA   ~for   accurate  metallicity   measurements.    The
astrometric  positions of  the  targets, taken  from  the 2MASS  Point
Source   Catalogue\footnote{http://www.ipac.caltech.edu/2mass/},  have
accuracies of $\sim0.1''$.

The  observations were  taken over  two  observing runs,  on 5--8  and
14--19 June 2008, with an  average seeing of $\sim1.5''$.  We obtained
3$\times$1200-second  exposures  per  fibre configuration,  to  obtain
S/N$\sim20-100$,  with several configurations  per cluster  (see Table
\ref{observed stars}  for actual numbers of spectra  obtained for each
cluster -- note  that for M68, 3 configurations  were obtained in June
2008,  and 5 further  configurations were  obtained during  a separate
service run in May  2009). In this study we also made  use of data for
M30  obtained in  a previous  observing run  in  2006 \cite[originally
published by][]{Kiss07}.  Flat fields, arc lamp exposures and $\sim$25
dedicated  sky fibres were  used for  data reduction  and calibration.
FeAr, CuAr, CuHe, CuNe and ThAr arc lamps were used to ensure accurate
wavelength  calibration.   Data  reduction  was  performed  using  the
{\tt2dfdr}
pipeline\footnote{http://www.aao.gov.au/2df/software.html\#2dfdr},
designed  specifically for  the  reduction of  AAOmega data.  \new{The
efficacy  of  the pipeline  has  been  checked  with a  comparison  of
individual stellar spectra.}

\subsection{Cluster Selection}\label{Cluster Selection}

We  aimed to  observe several  hundred stars  from each  of  four GCs,
namely M22  (NGC 6656), M30  (NGC 7099), M53  (NGC 5024) and  M68 (NGC
4590). These four targets were  selected by several criteria: they are
bright  ($M_V<-7$), nearby  ($D<30$\,kpc) and  have  radial velocities
$>1.5$\,$\sigma$  from the  peak  velocity given  by  the Besan\c  con
Galaxy   Model\footnote{http://www.obs-besancon.fr/modele/}   in  that
direction, for  ease of extracting  the cluster members  from Galactic
contaminants.  One cluster (M30)  was included specifically because it
was one of the targets of \citet{Scarpa07}, and it satisfied all other
selection criteria.  This allows a comparison with the results of that
study.

\begin{table}
\begin{center}
\caption{Clusters  selected  for  the  current survey,  based  on  the
  criteria  outlined in  the  text,  with the  total  number of  fibre
  configurations, and spectra, obtained  during the run. Note that the
  number  of  spectra quoted  here  is  the  total number  of  spectra
  obtained  before Galactic  contaminants  were removed.   For M68,  3
  configurations were observed on the  two observing runs in June 2008
  and 5  subsequent configurations were observed during  a service run
  in May 2009.}\label{observed stars}
\begin{tabular}{@{}ccc@{}}
  \hline
  Cluster & \# Fibre Configurations & Total \# Spectra\\
  \hline
  M22 & 10 & 3407\\
  M30 & 2 & 620\\
  M53 & 6 & 1727 \\
  M68 & 3+5 & 2650\\
  \hline
\end{tabular}
\end{center}
\end{table}

We produced  a Colour Magnitude  Diagram (CMD) for each  cluster using
data from the  2MASS Point Source Catalogue, and  the Red Giant Branch
(RGB)  was   identified  in  colour-magnitude   space.   Selection  of
individual  target  stars was  based  on  the  $J-K$ colours  and  $K$
magnitudes of the  RGB of the cluster to minimise  the number of stars
selected  from  the   Galactic  population.   Despite  this  selection
process, a number of Galactic contaminant stars were still expected in
the final sample and it  was therefore necessary to remove them before
analysis.  Section \ref{Cluster Membership} describes this process.

\subsubsection{Cluster Membership}\label{Cluster Membership}

To  select cluster  members,  it was  necessary  to determine  several
quantities for  each observed star. Atmospheric  parameters and radial
velocities were determined using  an iterative process, combining best
fits  to  the synthetic  spectra  from  the \citet{Munari05}  spectrum
library, with  $\chi^2$ fitting,  and cross-correlating this  best fit
model   with   the  observed   spectrum   to   calculate  the   radial
velocity. This approach is very  similar to that adopted by the Radial
Velocity Experiment  (RAVE) project \citep{Steinmetz06,Zwitter08}, and
is    based   on    the   same    synthetic   spectral    library   as
RAVE. \citet{Kiss07} outlined this method in detail.

For M22,  M30 and  M68, we used  four parameters to  identify members,
namely metallicity  ([m/H]), equivalent  width of the  calcium triplet
lines, radial velocity and the distance from the centre of the cluster
[terminating at the tidal  radius as quoted by \citet{Harris96}]; only
stars  that matched  using all  criteria  were judged  to be  members.
Figure  \ref{Cluster members  figure} shows  all the  stars  for which
spectra  were obtained,  highlighting those  determined to  be cluster
members.  Note  that spectra from observing runs  over 2006--2008 were
combined with the spectra of M22, M30 and M68 taken during the current
survey to ensure greater statistical significance of the results.

M53, being  closer to  the Galaxy in  both [m/H] and  radial velocity,
required a non-degenerate metallicity to determine membership. We used
the  method outlined by  \citet{Cole04} to  calculate [Fe/H]  from the
equivalent  widths  of  the  calcium  triplet  lines  for  each  star,
resulting in a  much cleaner selection for this  cluster. Further cuts
were made  on temperature ($T_{\rm  eff}<9000$\,K) and the  quality of
the  synthetic spectral  fit  to  produce the  final  sample used  for
analysis.  \new{Typical uncertainties in  the radial velocities of our
member   stars   are   $\sim3$\,km\,s$^{-1}$,   $\sim2$\,km\,s$^{-1}$,
$\sim3$\,km\,s$^{-1}$ and  $\sim2.5$\,km\,s$^{-1}$, for M22,  M30, M53
and  M68,  respectively.   All  stars  have  uncertainties  less  than
$6$\,km\,s$^{-1}$.}

The sample  consisted of 345,  194, 180 and  123 RGB member  stars for
M22, M30, M53  and M68 respectively.  Note that for  M30, the 168 Main
Sequence  Turn  Off  (MSTO)   and  sub-giant  branch  stars  from  the
\citet{Scarpa07} study  were also  included in our  analysis, bringing
the total number  of members for that cluster  to 362. The consequence
of mixing stellar types in our analysis of M30 is discussed in Section
\ref{veldisp}.

\begin{figure*}
  \includegraphics[angle=-90,width=0.245\textwidth]{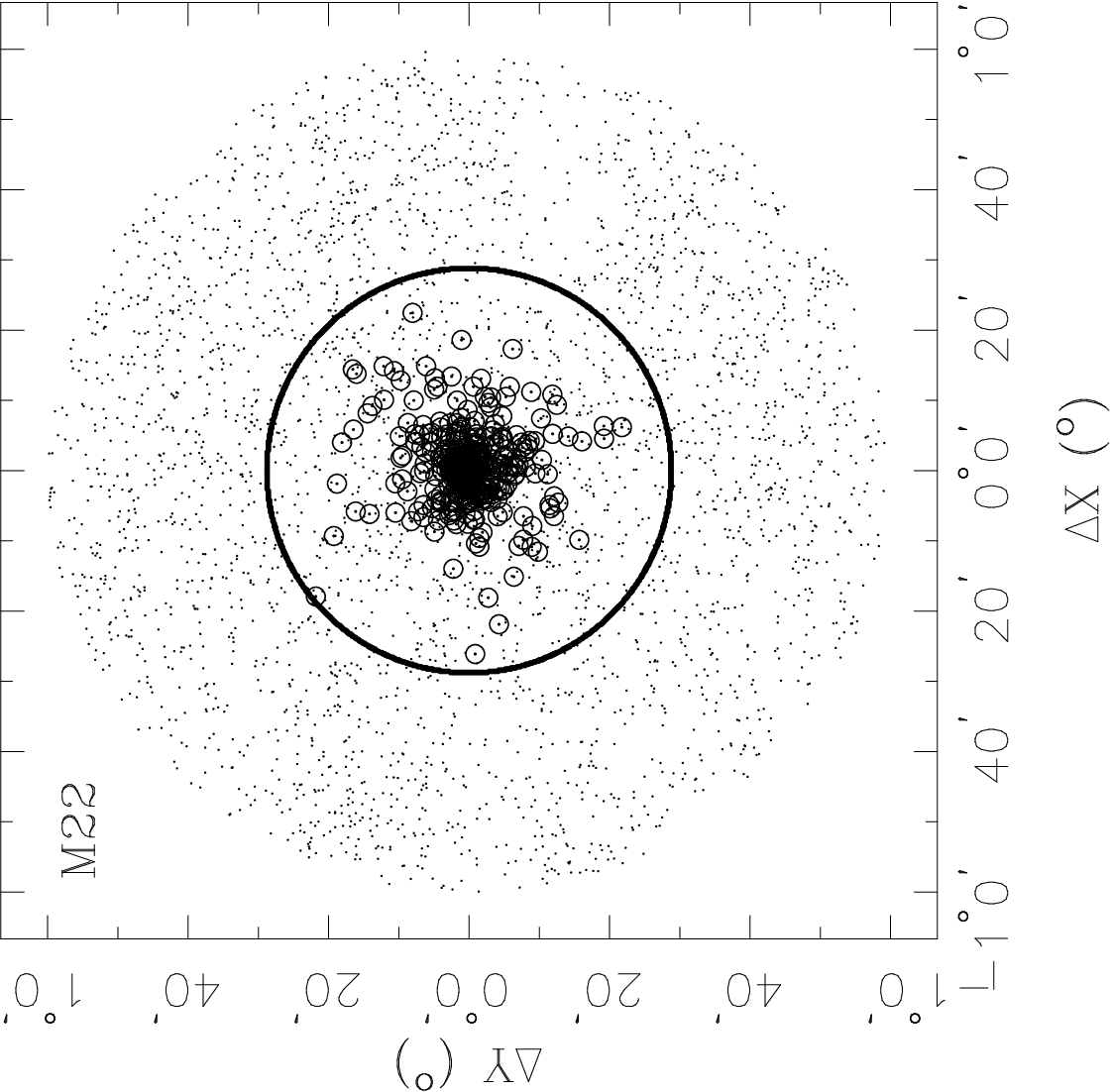}
  \includegraphics[angle=-90,width=0.245\textwidth]{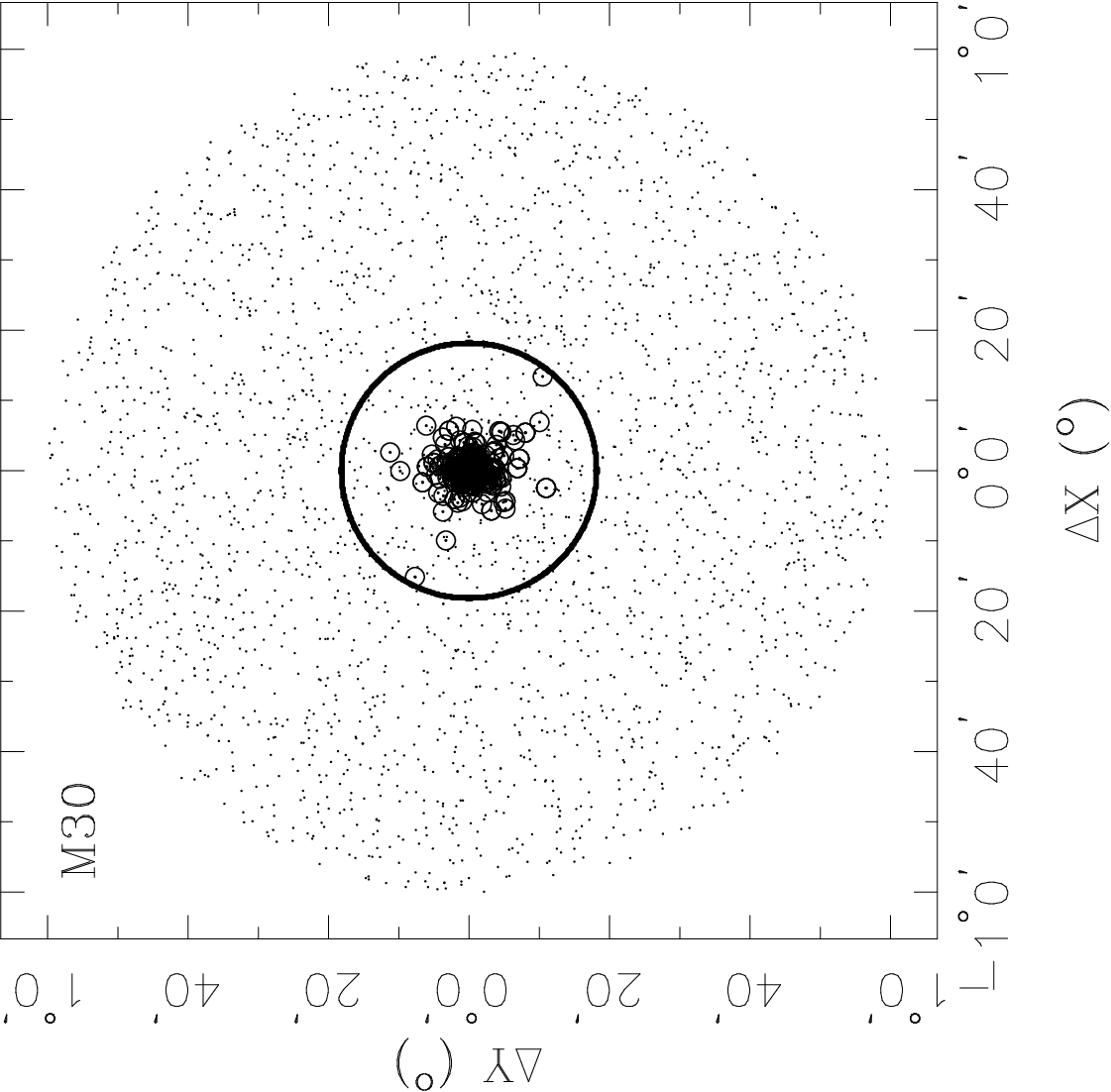}
  \includegraphics[angle=-90,width=0.245\textwidth]{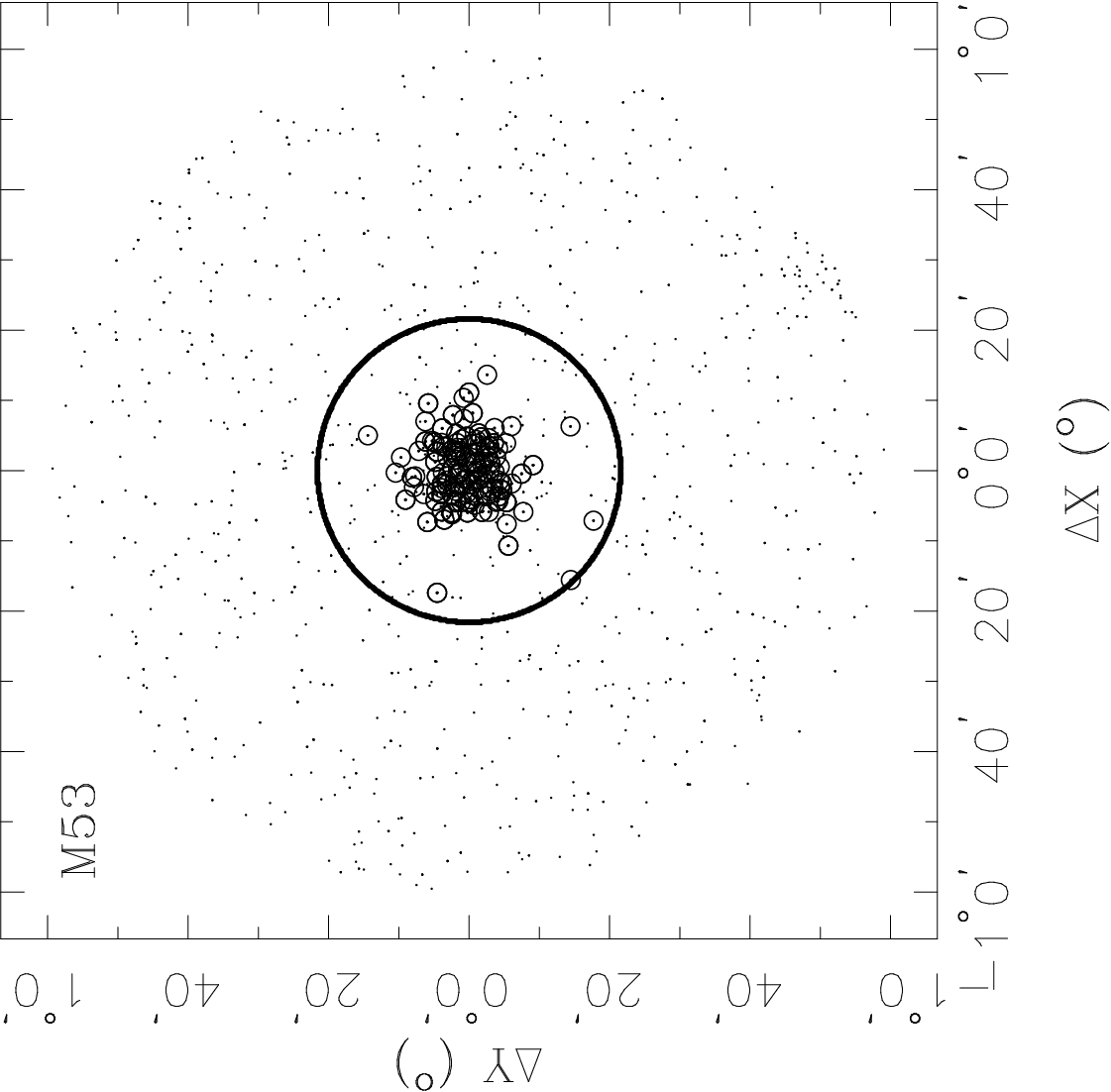}
  \includegraphics[angle=-90,width=0.245\textwidth]{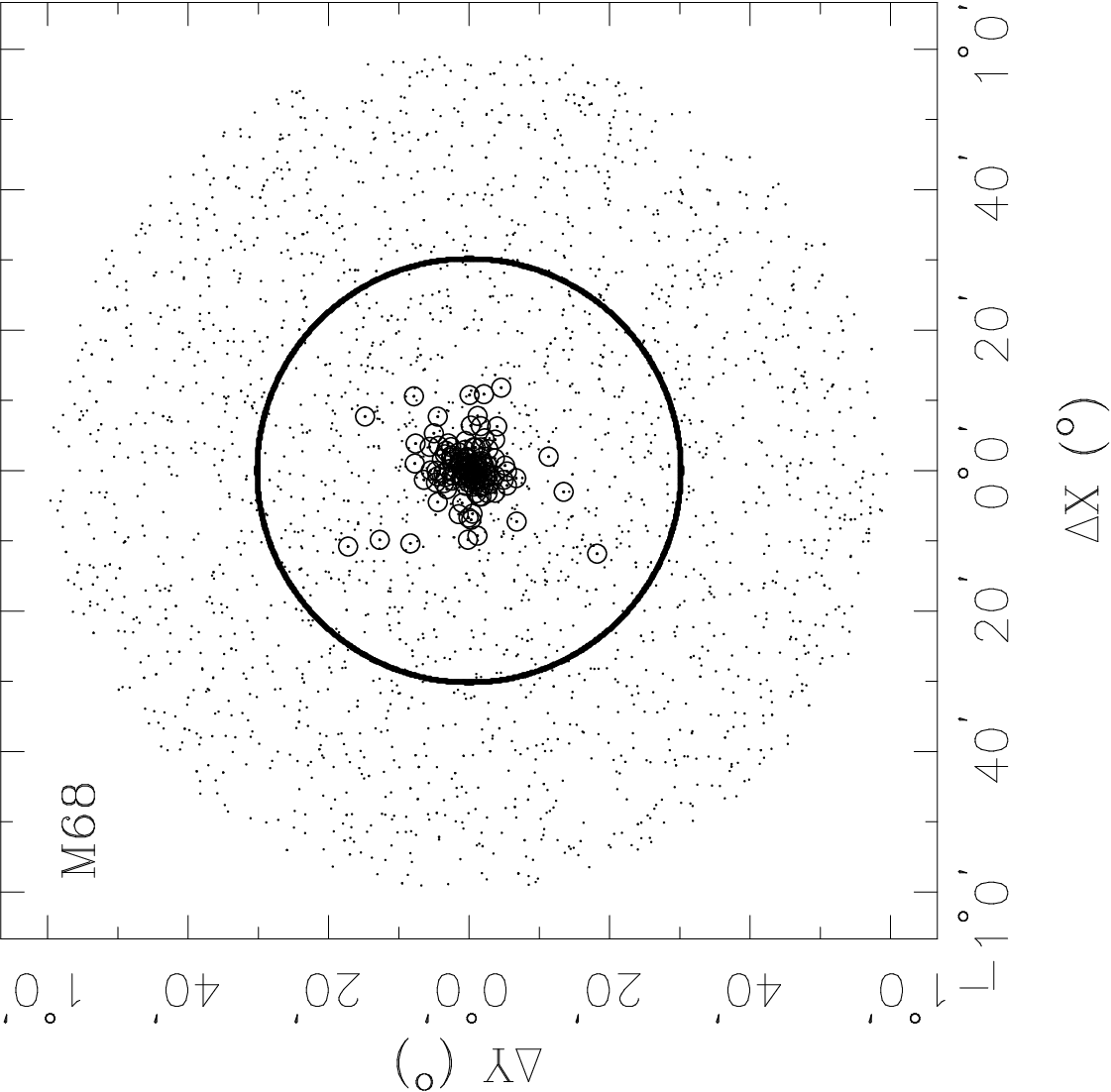}
  \caption{Members of each cluster used for analysis (circled dots),
    based  on  the  selection  method  described  in  the  text.   The
    uncircled dots  are the stars  which were observed  and determined
    not to be  cluster members.  The large circle  is the tidal radius
    of the cluster  from \citet{Harris96}. In each panel,  North is up
    and East is to the left.}
  \label{Cluster members figure}
\end{figure*}

\section{Results}

\subsection{Rotation}

\begin{figure*}
\begin{centering}
  \includegraphics[angle=-90,width=0.45\textwidth]{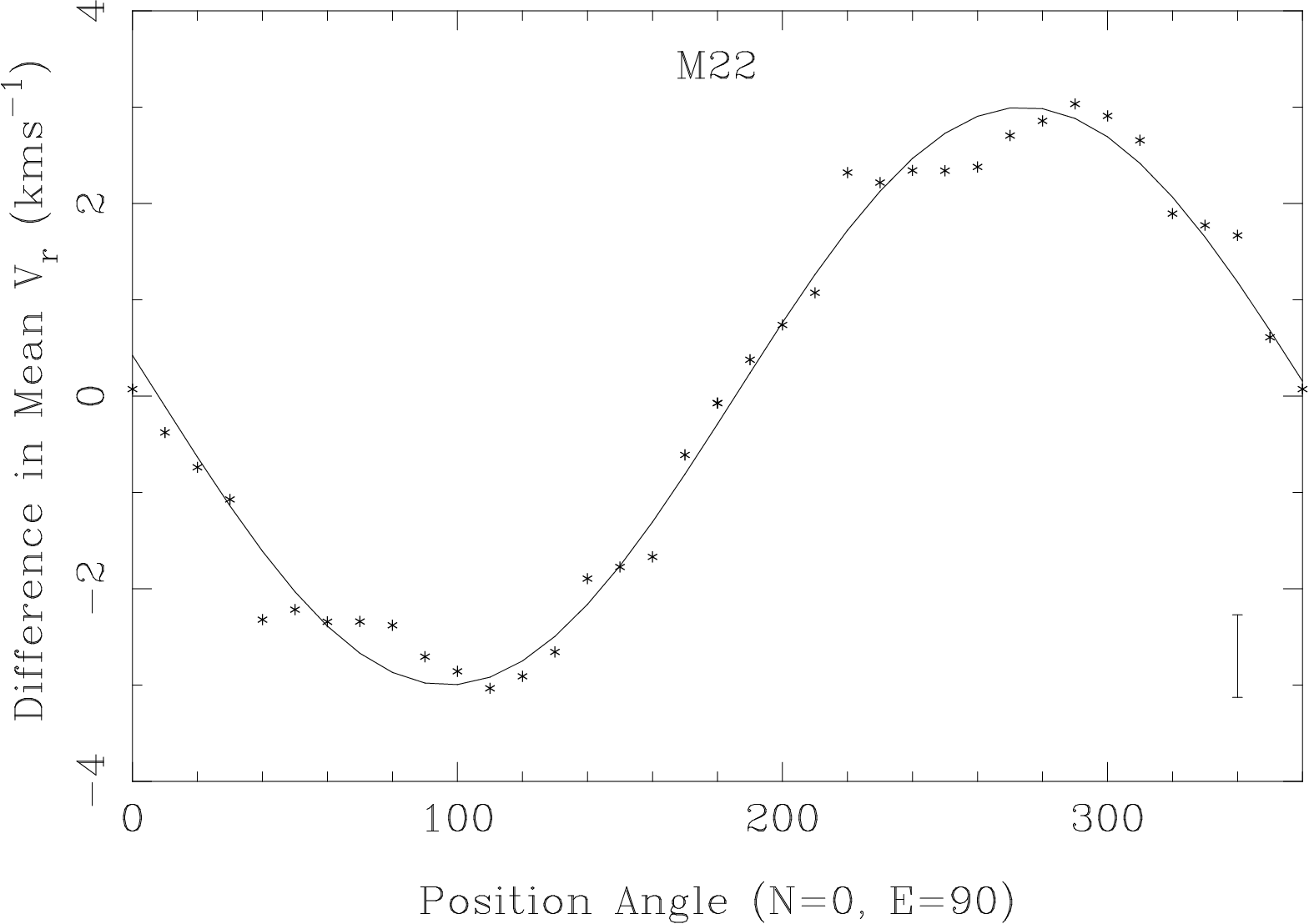}
  \includegraphics[angle=-90,width=0.45\textwidth]{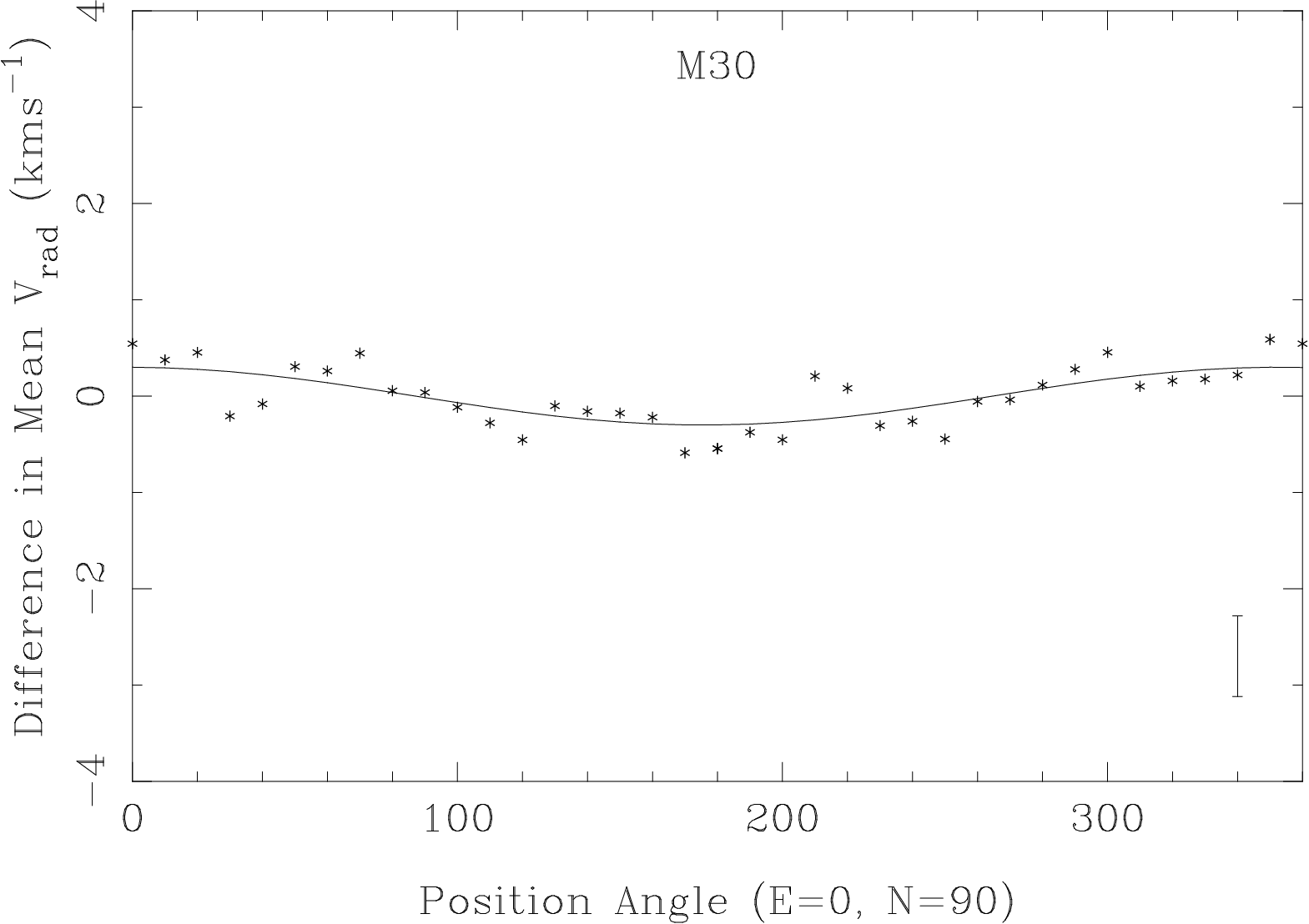}
  \includegraphics[angle=-90,width=0.45\textwidth]{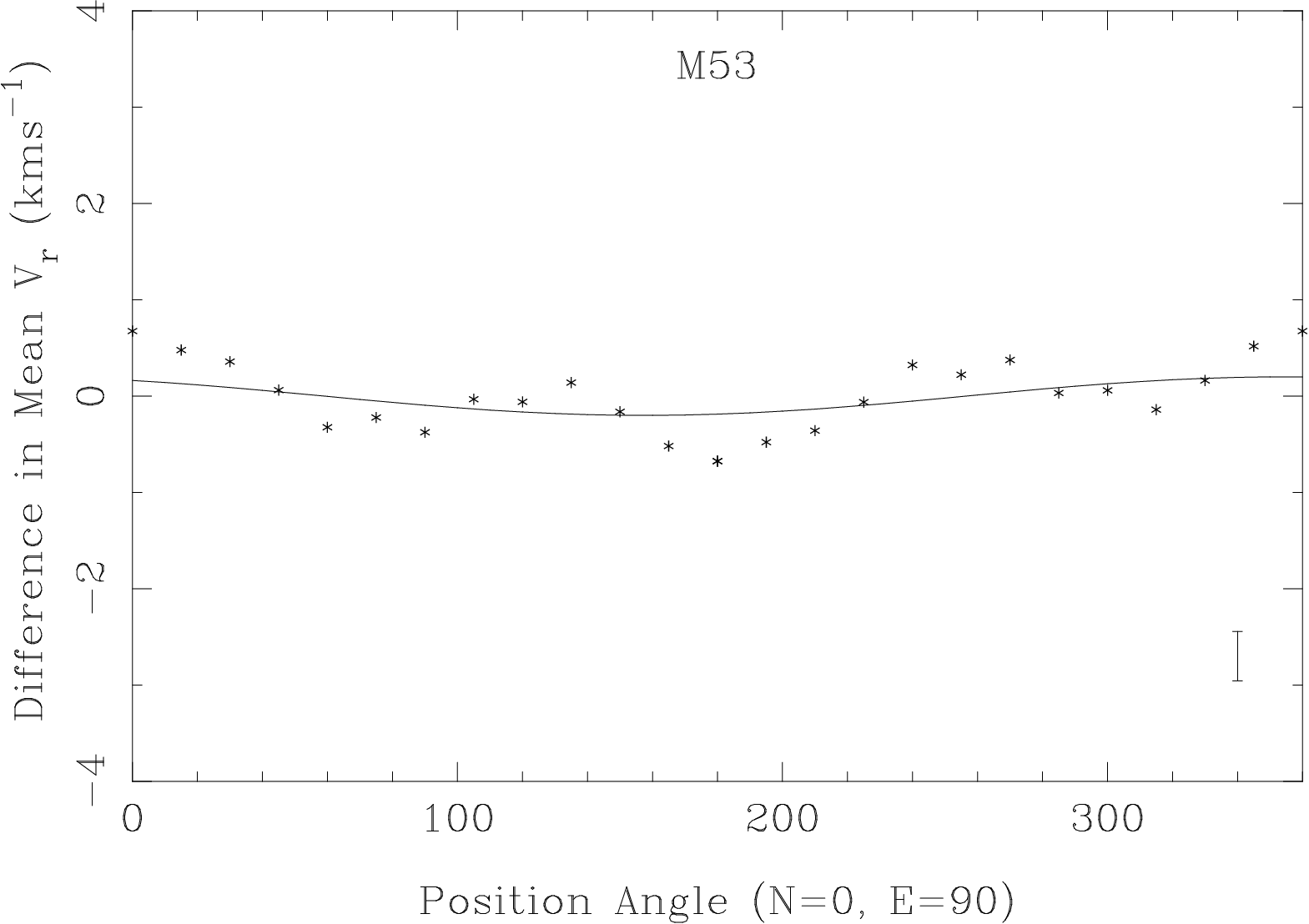}
  \includegraphics[angle=-90,width=0.45\textwidth]{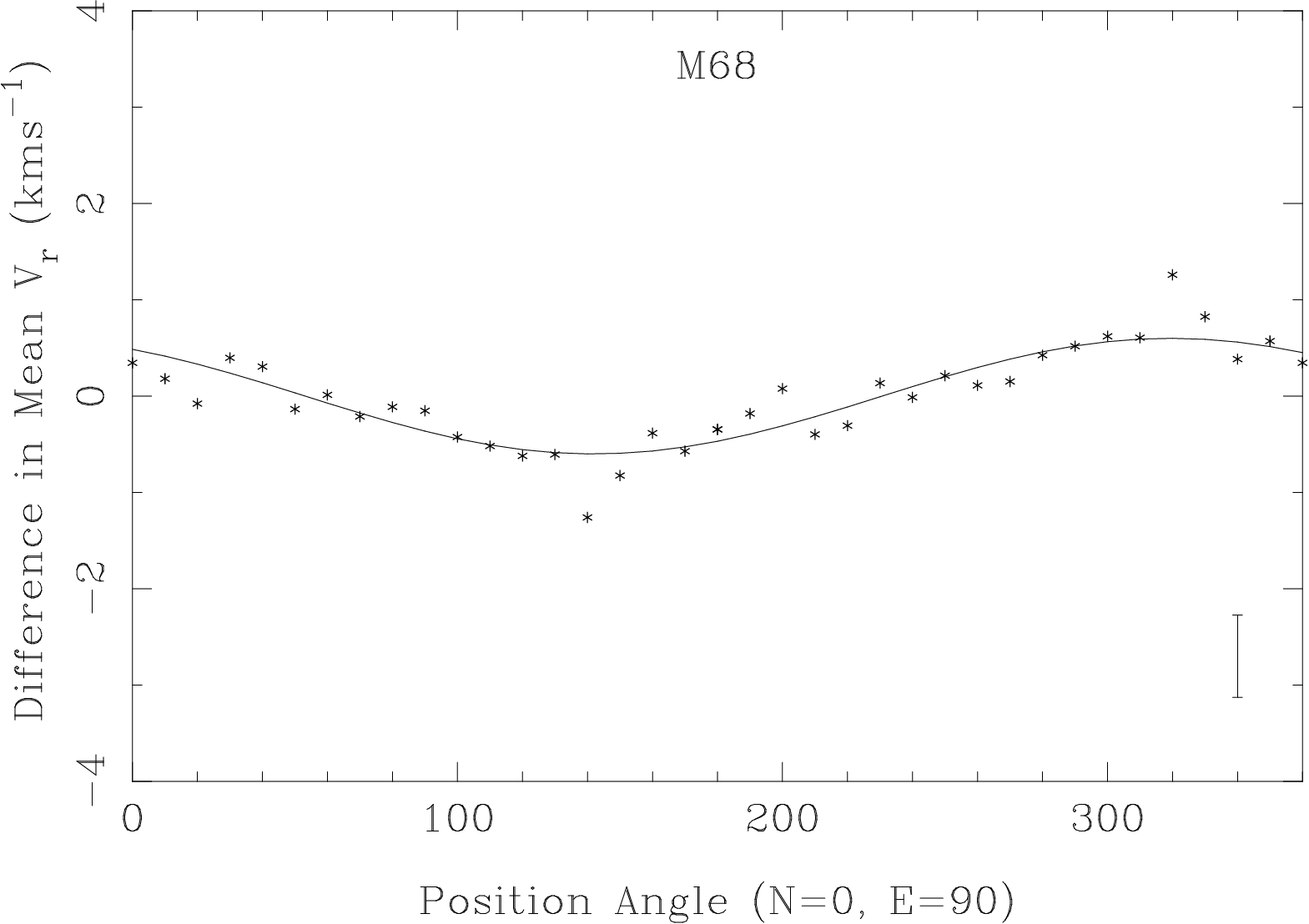}
  \caption{The rotation  of each cluster calculated  as the difference
    between the  mean velocities from  each side of the  cluster along
    equal  position  angles,  as  described  in the  text.   The  best
    $\chi^2$ fit sine function is overplotted, and a typical error bar
    is represented in the lower right of each panel.}
  \label{rotation}
\end{centering}
\end{figure*}

The  radial velocities  were used  to determine  the rotation  of each
cluster, assuming  an isothermal rotation.  The  rotation was measured
by dividing  the cluster in  half at a  given position angle  (PA) and
calculating  difference  between the  average  velocities  in the  two
halves.  This was performed over  PAs in steps of 10$^{\circ}$ for all
clusters, except  M53 was done  in 15$^\circ$ steps to  avoid aliasing
effects.   The resulting  data were  then  fit with  a sine  function.
Figure \ref{rotation} shows the measured rotation for each cluster and
the best-fit  sine function.   Note that \citet{Scarpa07}  detected no
rotation in M30 to a  level of 0.75\,km\,s{$^{-1}$} and we corroborate
their  result;  we detect  no  evidence for  rotation  to  a level  of
0.8\,km\,s$^{-1}$.    For    M53,   no   rotation    is   evident   to
0.5\,km\,s$^{-1}$, however, it  is clear that M22 is  rotating, with a
projected axis of rotation  approximately North-South.  M68 shows some
indication  of  rotation, although  additional  data  are required  to
confirm this.

\subsection{Velocity Dispersions}\label{veldisp}

To  determine  the  velocity  dispersions  of our  samples,  we  first
corrected all four clusters for rotation, and then binned the stars by
distance  from  the cluster  centre  (annuli), ensuring  approximately
equal numbers of  stars per bin. The mean  and standard deviation were
calculated for  the velocities in each bin.   Our velocity dispersions
were determined  using a Markov  Chain Monte Carlo  maximum likelihood
method (see \citealt{Gregory05} for  an overview of MCMC methodology),
which takes into account  the individual velocity uncertainties on the
stars and provides the velocity dispersion in each bin with associated
uncertainties.

\begin{figure*}
\begin{centering}
  \includegraphics[angle=-90,width=0.45\textwidth]{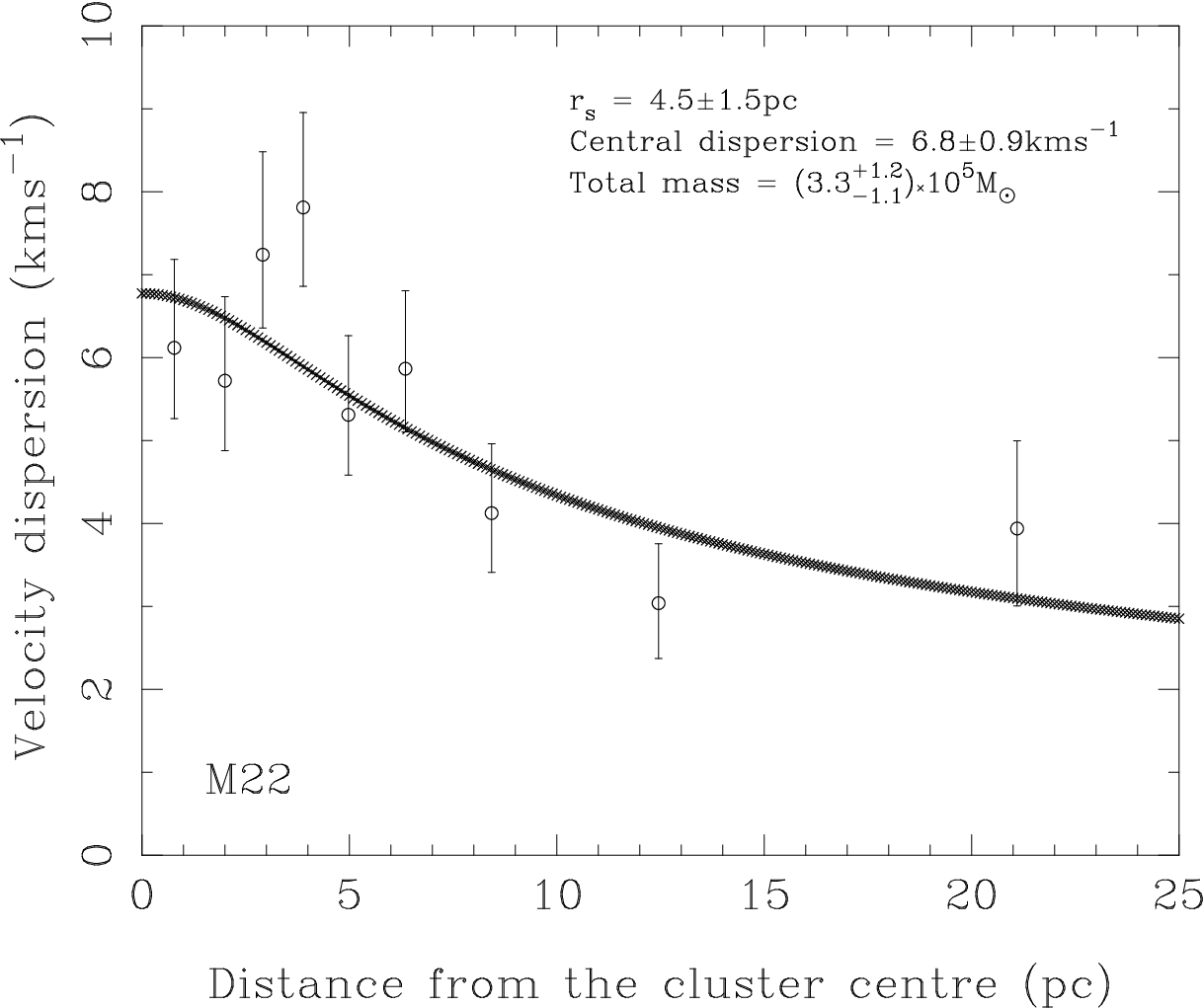}
  \includegraphics[angle=-90,width=0.45\textwidth]{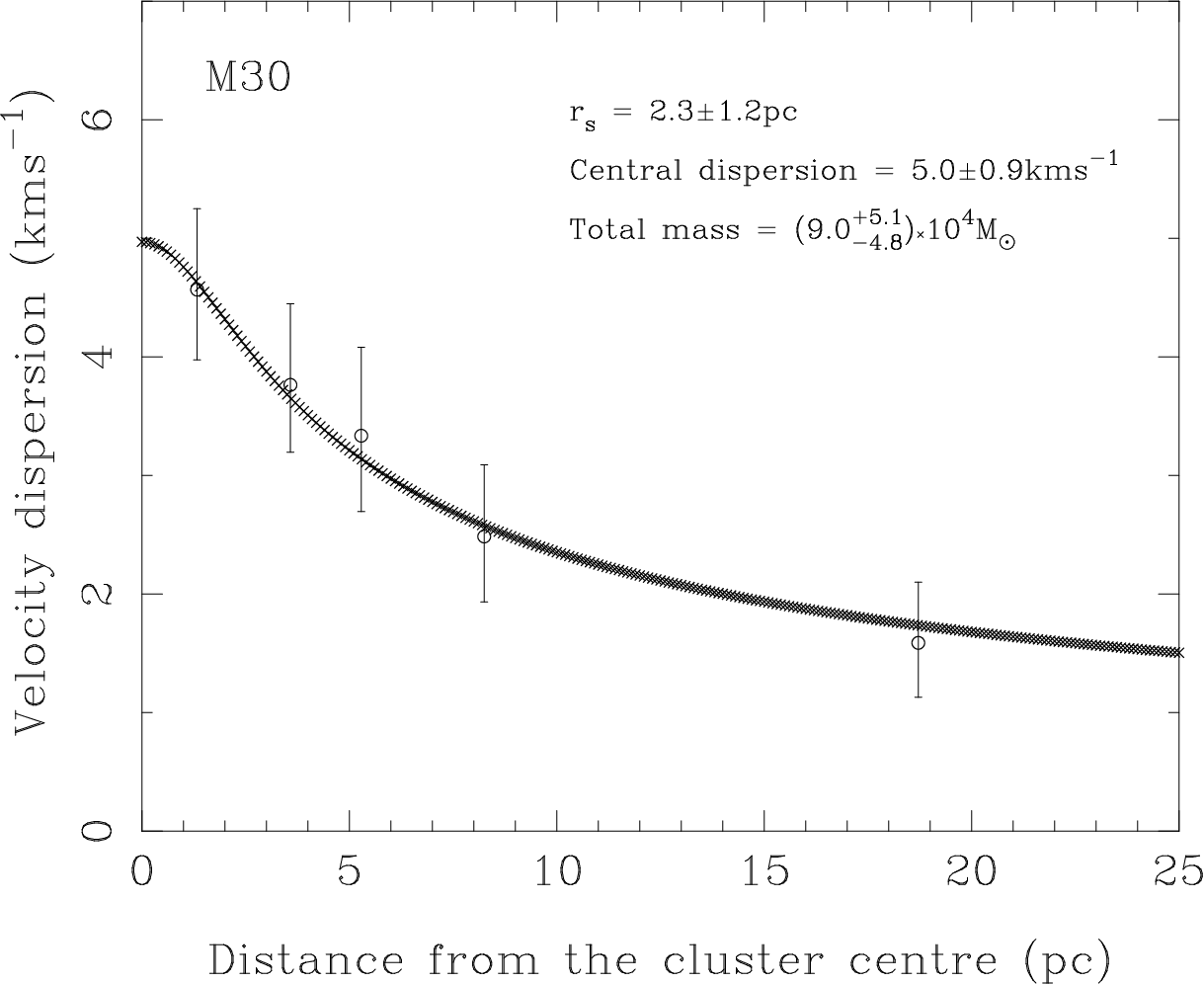}
  \includegraphics[angle=-90,width=0.45\textwidth]{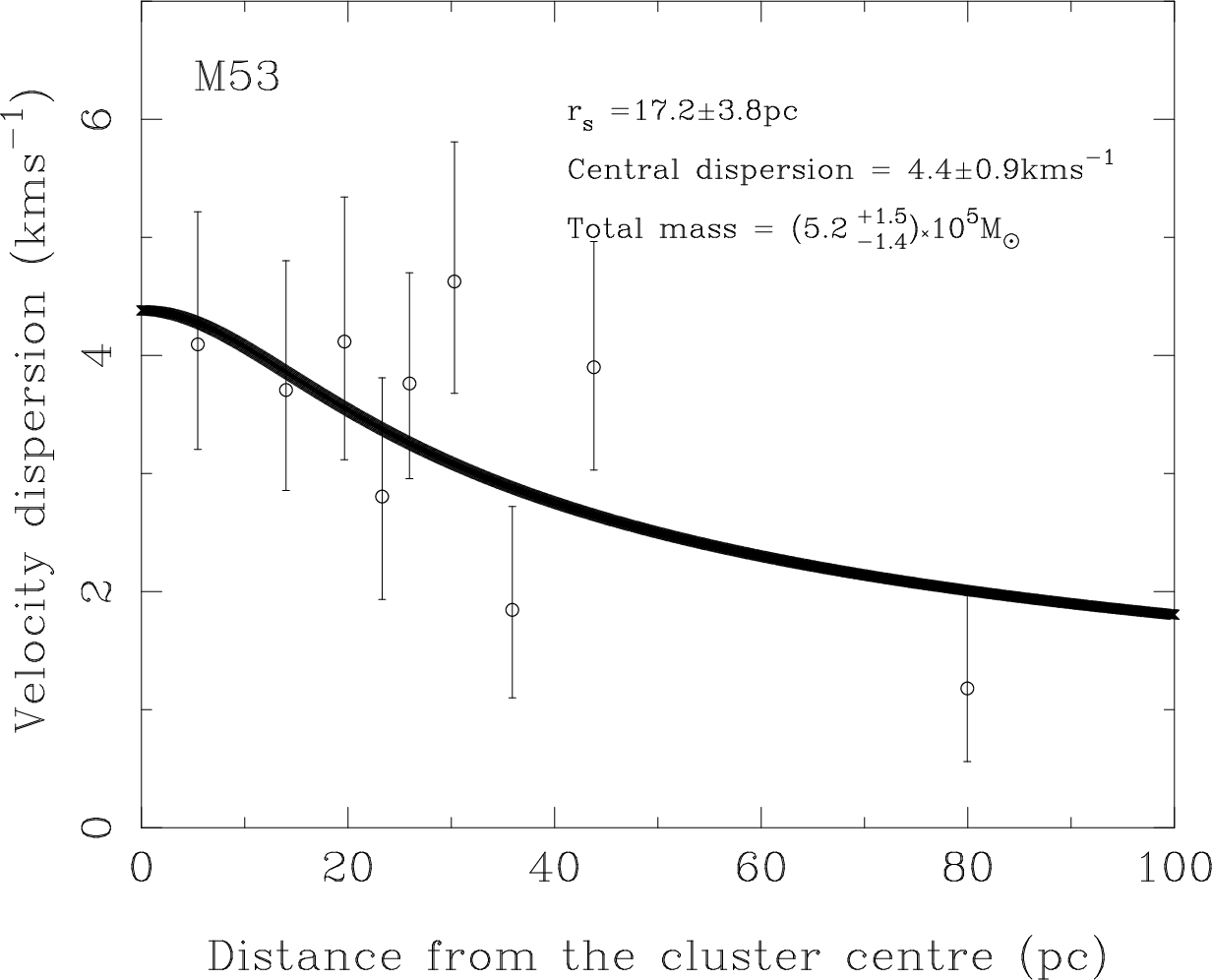}
  \includegraphics[angle=-90,width=0.45\textwidth]{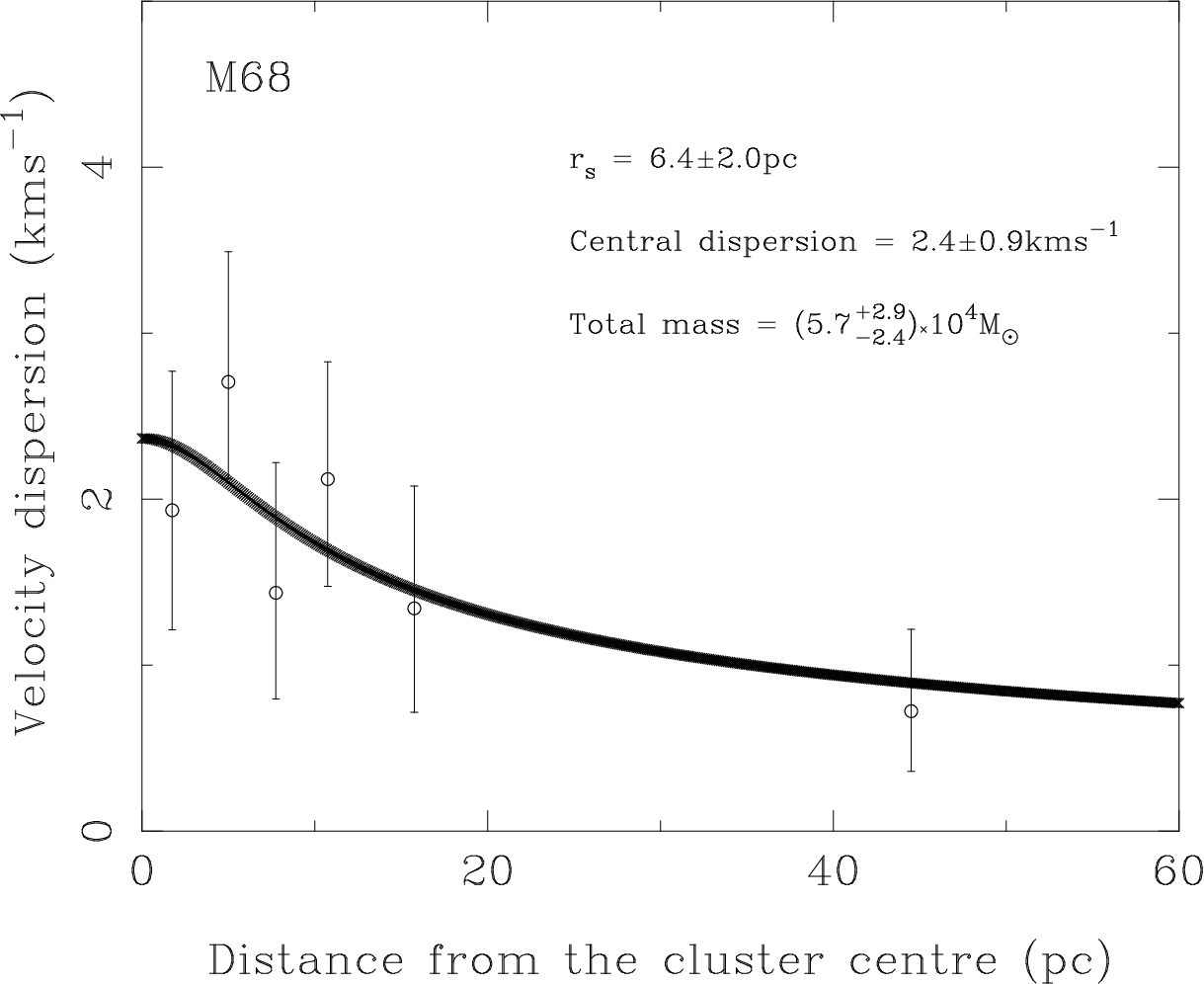}
  \caption{Velocity  dispersion  profiles of  each  cluster. The  best
  fitting  \citet{Plummer11}  model  is  overplotted, and  the  fitted
  parameters as well as the calculated total cluster mass is shown.}
  \label{dispersion profiles}
\end{centering}
\end{figure*}

The resulting velocity dispersion  profiles were then overplotted with
the   best-fitting  \cite{Plummer11}   model,  as   shown   in  Figure
\ref{dispersion profiles}.  \new{The data  points have been plotted at
the  mean radius  of the  stars in  each bin.   Because  fewer cluster
members exist  at larger  radii, the outermost  bin is  generally much
larger than  the inner bins. To  ensure our Plummer model  fits do not
exhibit an artificial slope in  the outskirts of the clusters, masking
any possible  flattening of the velocity dispersion  profiles, we also
performed fits  to the profiles with  the last bin  excluded. The fits
are  nearly  identical  for  M22,  M30  and  M68.   For  M53  a  small
discrepancy between  the fits was observed  due to the  scatter in the
velocity dispersions of the  inner bins, however, this discrepancy has
no impact on our overall results.  The reduced $\chi^2$ values for the
Plummer fits shown in  Figure \ref{dispersion profiles} are $\sim2.3$,
$\sim1.1$,  $\sim2.0$  and  $\sim1.5$  for  M22,  M30,  M53  and  M68,
respectively.}  We used two  parameters in the fitting process, namely
the  central   velocity  dispersion  and  the   scale  radius  ($r_s$;
containing half the cluster mass).  Due  to the nature of the model it
is  possible to  calculate the  total  cluster mass  from the  central
velocity dispersion  ($\sigma_0$) and $r_s$  (see \citealt{Dejonghe87}
for a detailed discussion of Plummer models and their application):

\[
M_{tot} = \frac{64\sigma_0^2r_s}{3\pi G}.
\]

\noindent  The  total  masses  are  shown  in  each  panel  of  Figure
\ref{dispersion profiles}.

We chose the Plummer model  to fit to our velocity dispersion profiles
because it is a  monotonically decreasing function and, therefore, any
flattening  of the  profiles would  be clearly  shown. The  cluster in
common between this study and that of \citet{Scarpa07}, M30, shows the
best Plummer  fit of  the four. \new{Note  that our  complete dataset,
which includes those data used by \citet{Scarpa07}, has been binned to
ensure  similar numbers  of stars  per  bin.  Because  our binning  is
different to that used by \citet{Scarpa07}, a data point-by-data point
comparison between the two studies  is not possible here.}  Within the
limits of the  model, and the uncertainties in  the data, the velocity
dispersion  does  not  exhibit  any  flattening at  large  radii,  and
therefore there is  no indication that either DM  or a modified theory
of gravity is required to explain the velocity dispersion of M30.  The
same case  can be made for  the other three clusters  we examined, and
the velocity  dispersion profiles, along with best  fit Plummer models
(and parameters) are presented in Figure~\ref{dispersion profiles}.

Different  stellar populations  in  GCs are  known  to have  differing
velocity  scatters  within   the  cluster  \cite[``velocity  jitter'';
see][for detailed  discussions]{Gunn79,Carney03}.  This jitter  is due
to instabilities in the atmospheres of luminous giant stars.  Since we
included the MSTO and  sub-giant stars from the \citet{Scarpa07} study
with our  RGB sample,  this jitter should  be evidenced as  a slightly
increased  velocity dispersion  from the  result  by \citet{Scarpa07}.
This  increase in  dispersion will  be  smaller, however,  than if  we
produced our velocity dispersions from our RGB stars alone.  Comparing
our dispersion profiles for M30  with those of \cite{Scarpa07}, we see
some evidence for  this increase in the inner  regions of the cluster,
although the difference is  well within the uncertainties. No increase
in dispersion  is seen  in the outer  parts; \citet{Scarpa07}  quote a
dispersion of $\sim$2.25\,km\,s$^{-1}$ at  a radius of 25\,pc, whereas
our  data  show  a   dispersion  of  $\sim$1.6\,km\,s$^{-1}$  at  this
distance, although this discrepancy is still within the uncertainties.

The  main  result from  the  \citet{Scarpa07}  study  was an  apparent
flattening of  the velocity dispersion  profiles of five of  their six
clusters, indicating a DM component,  or a need for a modified gravity
theory  to  explain  their  results.   \citet{Moffat08}  overlaid  the
dispersion profiles of \citet{Scarpa07}  with a MOG dispersion profile
and showed  that the  flattening of the  profiles was  consistent with
Newtonian gravity, although  the authors did omit one  data point from
two of the clusters  analysed.  The \citet{Moffat08} MOG model differs
from Newtonian gravity for very massive objects (e.g.  galaxy clusters
and elliptical galaxies) but becomes  Newtonian for masses below a few
times $10^6$\,M$_\odot$, which is the  mass range of GCs.  The MOG fit
proves to  be an  equivalently good fit  to these data,  indicating no
deviation  from Newtonian gravity  in these  clusters, and  a Gaussian
with   a   flat   tail   has  no   physical   significance.    Because
\citet{Scarpa07} overlaid their data with a Gaussian with a flat tail,
and \citet{Moffat08}  overlaid the same  data with a MOG  model (which
becomes Newtonian  at low  mass), and both  are a similarly  good fit,
neither DM nor  a modified version of gravity  are required to explain
their data, and our study confirms this.

\subsection{Systemic Velocities}

In addition to the velocity dispersions, the systemic velocity of each
cluster was  measured (using the  Monte Carlo method  described above)
and   compared  to   similar   data  in   \cite{Harris96},  the   most
comprehensive  catalogue   of  GC  parameters.    This  comparison  is
summarised in Table  \ref{velocity comparisons}.  Note especially M53,
for which  our calculated velocity is significantly  different to that
of \cite{Harris96},  with the discrepancy well outside  the error bars
of both measurements.   Since \cite{Harris96} quotes systemic velocity
data for M53 from \cite{Webbink81},  who examined a total of 12 stars,
our  systemic  velocity calculation  based  on  180  stars is  a  more
reliable measurement for this cluster.

\begin{table}
\begin{center}
\caption{Comparisons between the radial  velocities of each cluster in
  the    \citet{Harris96}    catalogue    and    those    from    this
  survey.}\label{velocity comparisons}
\begin{tabular}{@{}cccc@{}}
  \hline
  Cluster & V$_r$ \citep{Harris96} & V$_r$ (this paper)\\
  \hline
  M22 & -148.9$\pm0.4$ & -144.86$\pm0.34$\\
  M30 & -181.9$\pm0.5$ & -184.40$\pm0.20$\\
  M53 & -79.1$\pm4.1$  & -62.80$\pm0.31$\\
  M68 & -94.3$\pm0.4$  & -94.93$\pm0.26$\\
  \hline
\end{tabular}
\end{center}
\end{table}

\subsection{Mass-to-Light Ratio}

One  important indicator  of large  quantities  of DM  in a  dynamical
system  is  a large  mass-to-light  ratio  (i.e.  M/L$_{\rm  V}\gg$1).
Furthermore, this DM will cause the stars to have large accelerations,
and therefore inherently higher maximal velocities, and hence a larger
velocity dispersion.   An obvious  way to determine  whether or  not a
pressure-supported object  like a  GC is dark  matter dominated  is to
measure  its  M/L$_{\rm  V}$.   Of  course, this  requires  a  surface
brightness profile, for which  we have used those by \citet{Trager95}.
These profiles were converted to solar luminosities per square parsec.
The   density   profile  was   then   calculated  using   \cite[again,
see][]{Dejonghe87}:

\[
\rho(r) = \frac{M_{tot}}{\pi}\frac{r_s^2}{(r_s^2 + r^2)^2}.
\]

\begin{figure*}
\begin{centering}
  \includegraphics[angle=-90,width=0.45\textwidth]{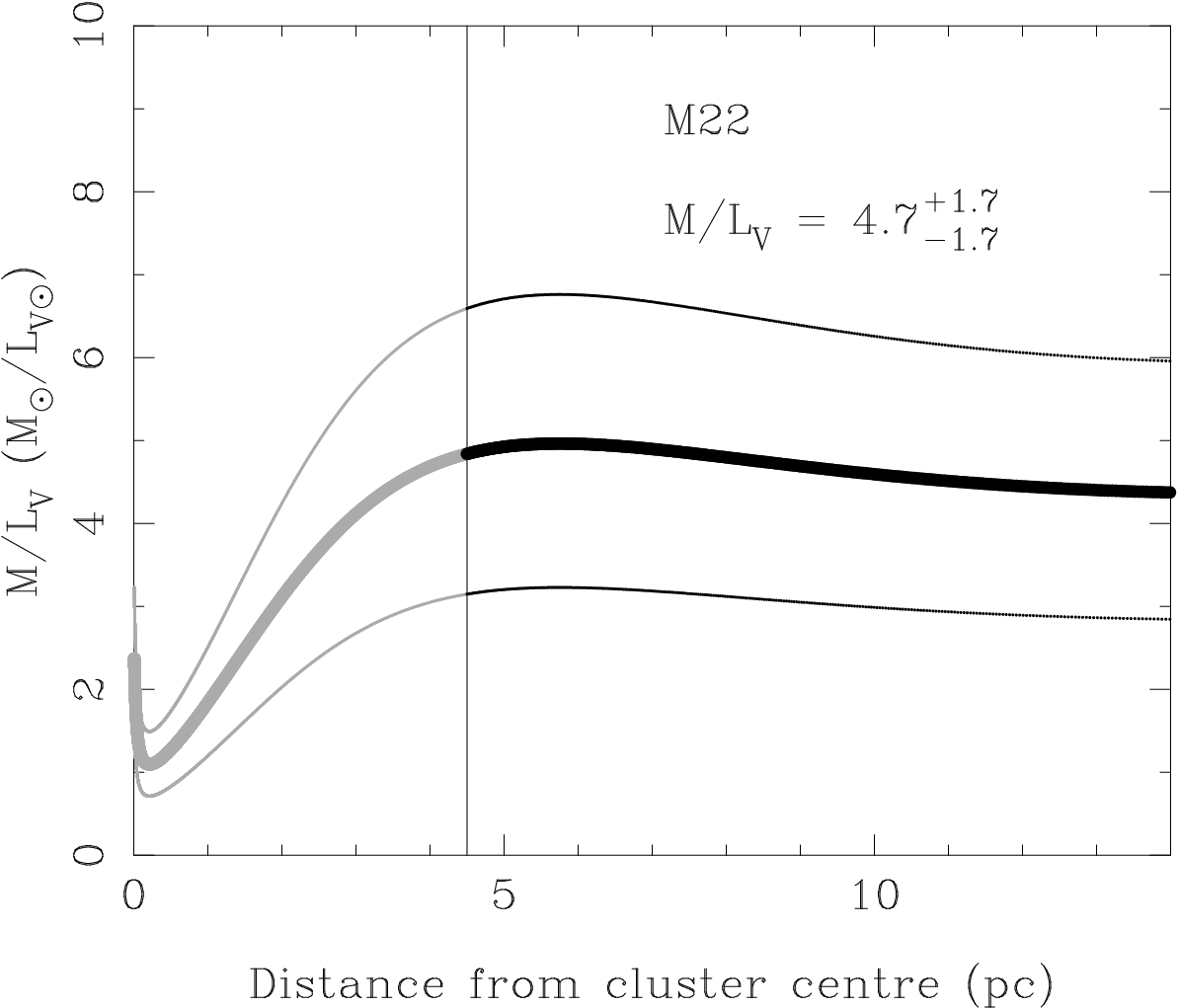}
  \includegraphics[angle=-90,width=0.45\textwidth]{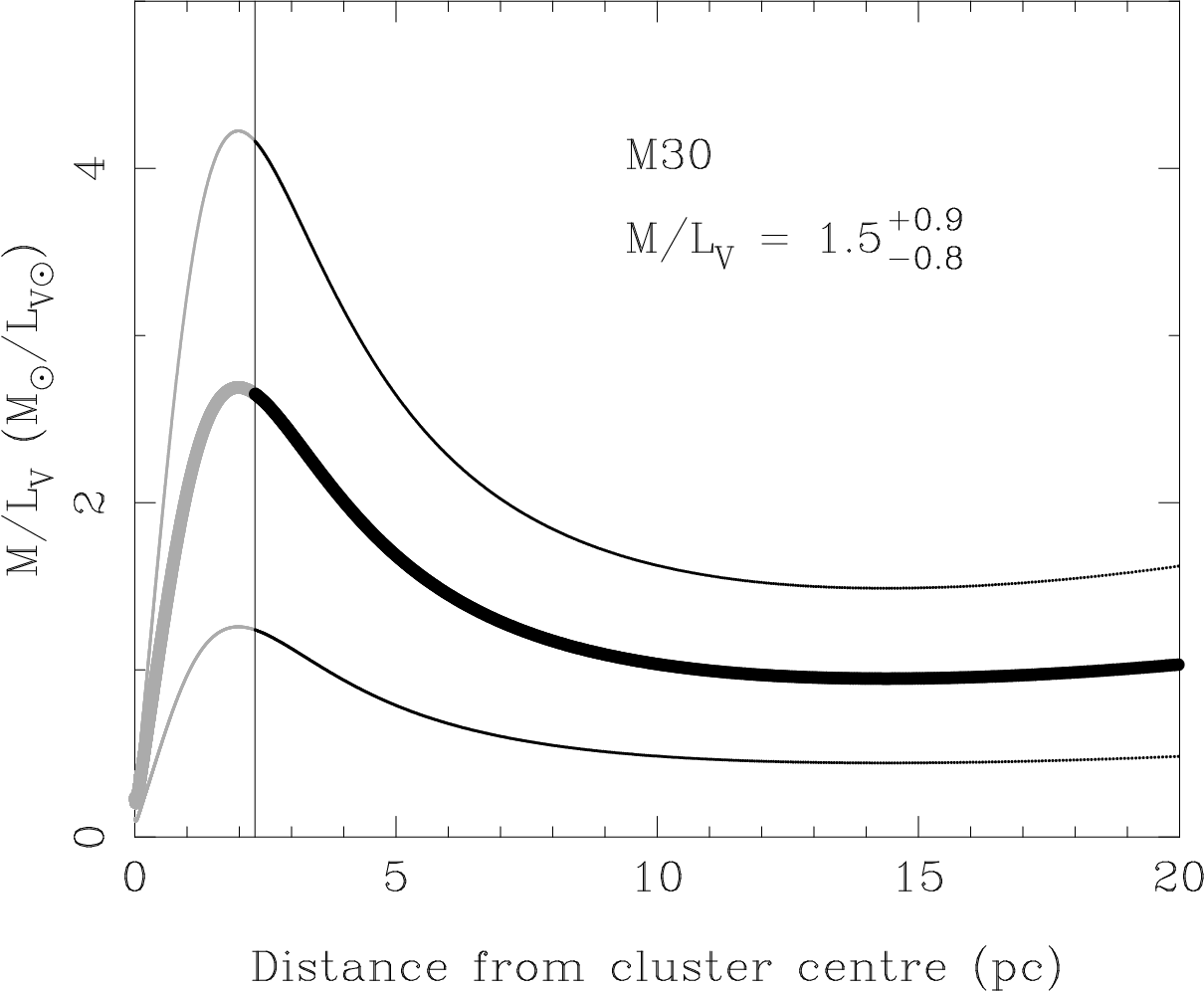}
  \includegraphics[angle=-90,width=0.45\textwidth]{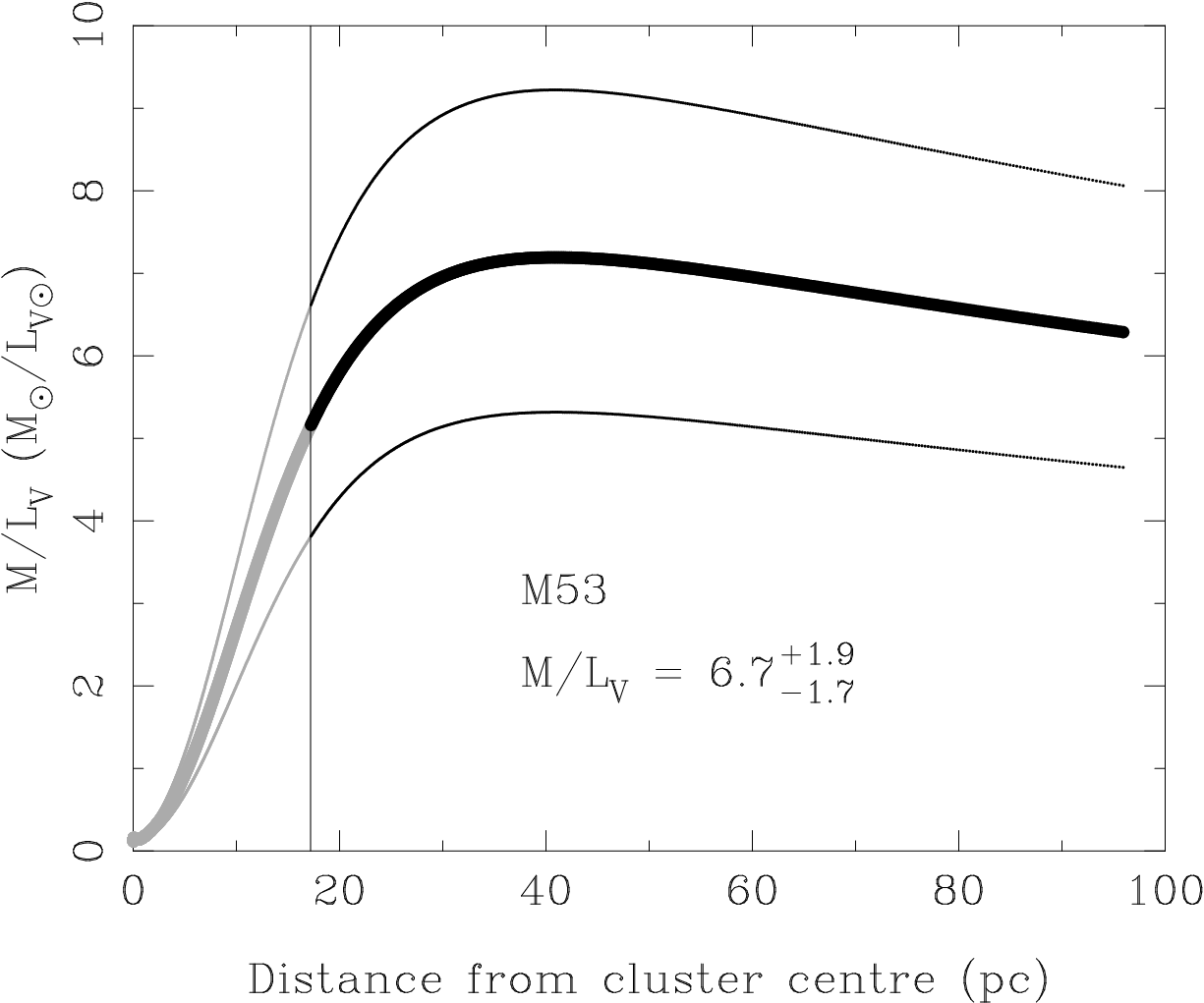}
  \includegraphics[angle=-90,width=0.45\textwidth]{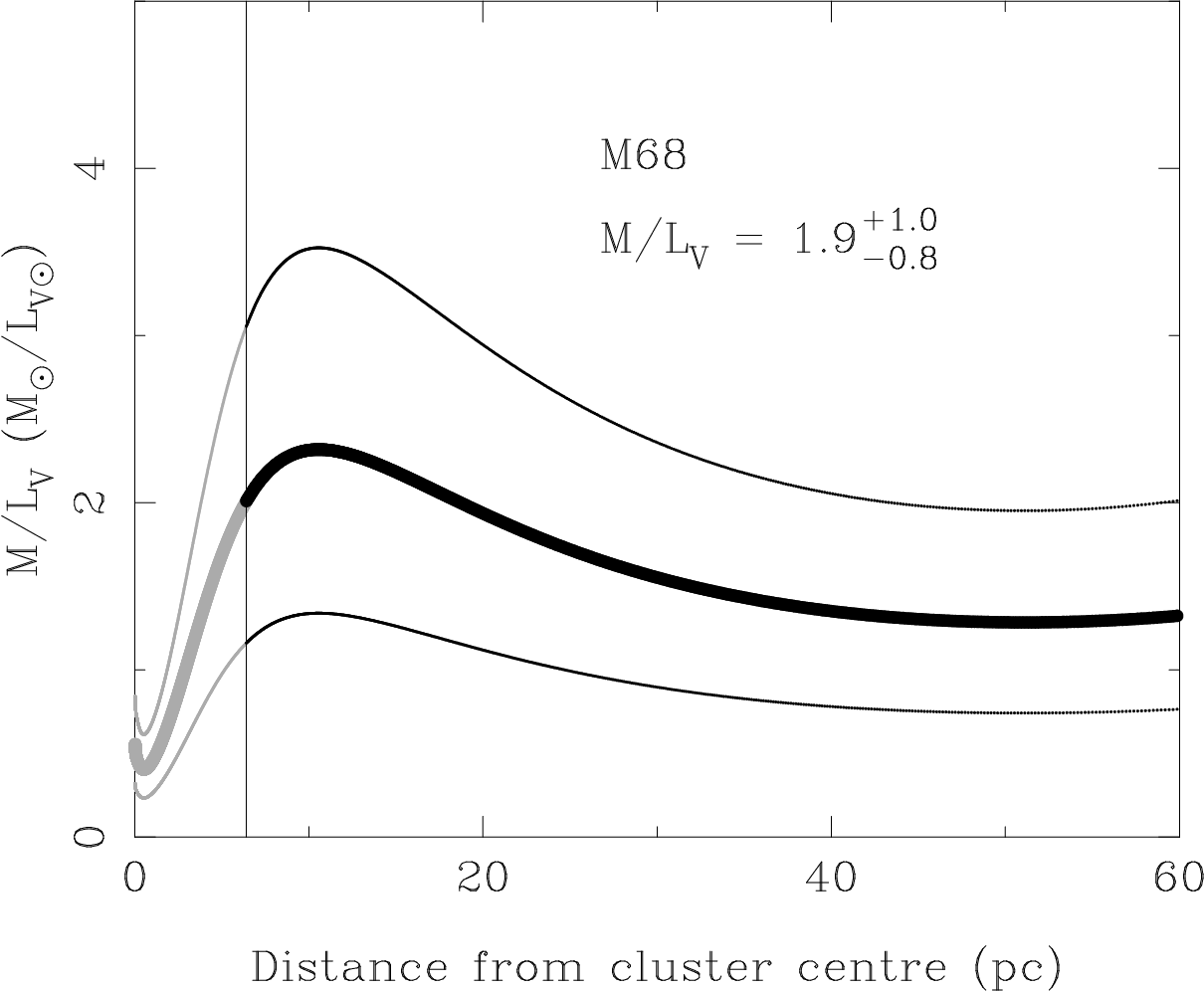}
  \caption{Mass-to-light profiles for each cluster (thick curve).  The
    thin  curves   show  the  uncertainties  on   the  M/L$_{\rm  V}$,
    calculated  via the  $1$\,$\sigma$  difference of  $r_s$ from  its
    calculated value.  The vertical line indicates the value of $r_s$.
    The M/L$_{\rm V}$ value given is only calculated for radii greater
    than $r_s$ because the mass  estimates are highly uncertain in the
    cores  of  GCs,   particularly  if  they  are  post-core-collapsed
    clusters.}
  \label{M/L}
\end{centering}
\end{figure*}

\noindent The  M/L$_{\rm V}$ profiles  are shown in  Figure \ref{M/L},
along with the M/L$_{\rm V}$ ratios. The vertical line shows the value
of $r_s$.  Because the central regions of GCs are highly concentrated,
particularly  in  post-core-collapsed  clusters,  it is  difficult  to
accurately determine the  mass in the core.  The  M/L$_{\rm V}$ ratios
have, therefore,  only been calculated  for radii greater  than $r_s$.
\new{\cite{Evstigneeva07}  showed  that  ultra-compact dwarf  galaxies
(UCDs) follow  the same luminosity -- velocity  dispersion relation as
old GCs.   Since it  has also  been shown that  UCDs with  a dynamical
M/L$_{\rm V}$  up to  5 do  not require dark  matter to  explain these
mass-to-light ratios \cite[e.g.][]{Hasegan05,Evstigneeva07}, we see no
requirement for  any of our clusters,  apart from M53,  to contain any
dark matter  component.  \cite{Hasegan05} determined  that a M/L$_{\rm
V}>6$ for  UCDs may  indicate some dark  matter content, and  since we
estimate M/L$_{\rm  V}\approx6.7$ for M53,  this may indicate  a small
dark matter component in this cluster.  Note that if we take the lower
limit of our uncertainties for the mass-to-light ratio of M53 we reach
the regime where dark matter is not required, and none of our clusters
show M/L$_{\rm V}\gg1$, indicating DM is not dominant.}

Note that  we have produced  M/L$_{\rm V}$ {\it profiles}  rather than
simply deriving a single M/L$_{\rm V}$ value from the central velocity
dispersion   and  central   surface  brightness.    We   suggest  that
determining M/L$_{\rm V}$ profiles  is preferable because it describes
the M/L$_{\rm V}$ of the  entire cluster outside the core, rather than
just  its  core.   This  is  especially  true  for  post-core-collapse
clusters, such as M30, in  which the core masses are highly uncertain.
Furthermore, M30  is the only  cluster in our  sample known to  have a
collapsed  core  \cite[e.g.][]{Trager95}.   This  cluster also  has  a
maximum in its M/L$_{\rm V}$  profile at about the scale radius $r_s$,
which  may be due  to mass  segregation.  The  most massive  stars are
known  to fall toward  the core  during the  evolution of  the cluster
\cite[see][for   a   review   of   mass   segregation   processes   in
GCs]{Spitzer85}.   The   somewhat  less  pronounced   maximum  in  the
M/L$_{\rm  V}$ profile  of M68  may be  the first  evidence  that this
cluster  is currently undergoing  core collapse.   It has  been argued
that  a central  concentration  parameter $c\approx2$  may indicate  a
collapsed   core    in   certain    GCs,   and   for    M68   $c=1.91$
\citep{vandenBergh95}.  \new{Note that, despite our calculation of the
M/L$_{\rm V}$  ratio being restricted to $R>r_s$,  the profiles should
be  accurate  for  $R>r_c$,  the  core radius,  or  $R\gtrsim  r_s/2$.
Therefore, we argue that the peak in M/L$_{\rm V}$ is real, although a
larger sample of cluster members  for $R<r_s$ will be required to test
this  and, therefore,  to show  whether M68  is truly  undergoing core
collapse.}

\section{Conclusions}

We  have  studied four  GCs  to  determine  their velocity  dispersion
profiles, and found that neither  DM nor modified gravity are required
to reconcile  the kinematic  properties of these  particular clusters.
\new{M53 may  contain a small  DM component, indicated by  a M/L$_{\rm
V}>6$,  but  it  is  not  dominant.   Within  the  uncertainties,  the
M/L$_{\rm V}$ ratio of M53 is  still consistent with little, or no, DM
content.}  The dynamics  of all four clusters are  well described by a
purely   analytic  \citet{Plummer11}   model,  which   indicates  that
Newtonian gravity adequately describes their velocity dispersions, and
shows      no     breakdown      of      Newtonian     gravity      at
$a_0\approx1.2\times10^{-10}$\,m\,s$^{-2}$,  as  has  been claimed  in
previous studies.

Furthermore, the Plummer  model was used to determine  the total mass,
scale radius (the radius at which half the mass is contained), and the
M/L$_{\rm  V}$ profile for  each cluster.   We find  that none  of our
clusters  have M/L$_{\rm V}\gg1$,  another indication  that DM  is not
dominant.  Within the uncertainties,  our estimated cluster masses all
match those  in the literature \cite[e.g.][]{Meziane96},  and the same
is true for the  M/L$_{\rm V}$ ratios \cite[e.g.][]{Pryor93}.  We have
produced M/L$_{\rm  V}$ {\it profiles},  rather than quoting  a single
value  based on the  central velocity  dispersion and  central surface
brightness.   This  method  is   preferred  because  it  describes  the
M/L$_{\rm V}$ of  the entire cluster, rather than  only its core; this
is particularly  important for  post-core-collapsed GCs for  which the
central mass is highly uncertain.

Because the only known of our  clusters to have a collapsed core (M30)
shows a peak in its M/L$_{\rm V}$ profile at approximately $r_s$, this
may be an indication of the resulting mass segregation. If this is the
case,  then  we may  have  the first  evidence  that  M68, which  also
exhibits  a  M/L$_{\rm   V}$  peak  near  $r_s$  and   has  a  central
concentration of  1.91 \citep{vandenBergh95}, is  currently undergoing
core collapse.  Further observations are required, both photometric and
kinematic, to confirm this.

Another important result from this  study is the measured rotations of
our  clusters.  Of  the four  clusters studied  here, one  exhibits an
obvious rotation,  namely M22, with an axis  of rotation approximately
North-South.   M68 may  be rotating  as well  and additional  data are
required to confirm this.  M53 and M30 do not show any rotation to the
level of 0.5km\,s\,$^{-1}$ and 0.8km\,s\,$^{-1}$ respectively.

While these results are strongly  indicative of the current picture of
globular clusters  being dark matter  poor, and their dynamics  can be
explained by  standard Newtonian theory, we  are currently undertaking
robust dynamical modelling of GC systems to fully address the question
of  the   influence  of   non-Newtonian  physics  in   explaining  our
observations.


\section{Acknowledgments}

This project  has been supported by  the Anglo-Australian Observatory,
the Australian Research Council and the Hungarian OTKA grant K76816.

\bsp

\label{lastpage}

\end{document}